\documentstyle[aps,floats,epsf]{revtex}

\begin{document}
\draft
\twocolumn[\hsize\textwidth\columnwidth\hsize\csname@twocolumnfalse%
\endcsname

\title{Threshold Resummation in Momentum Space from Effective Field Theory}

\author{Thomas Becher$^a$ and Matthias Neubert$^b$}

\address{$^a$\,Fermi National Accelerator Laboratory, P.O. Box 500, Batavia, 
IL 60510, U.S.A.\\
$^b$\,Institute for High-Energy Phenomenology, Laboratory for 
Elementary-Particle Physics,\\ 
Cornell University, Ithaca, NY 14853, U.S.A.}
\maketitle

\begin{abstract}
Methods from soft-collinear effective theory are used to perform the threshold 
resummation of Sudakov logarithms for the deep-inelastic structure function 
$F_2(x,Q^2)$ in the endpoint region $x\to 1$ directly in momentum space. An 
explicit all-order formula is derived, which expresses the short-distance 
coefficient function $C$ in the convolution $F_2=C\otimes\phi_q$ in terms of 
Wilson coefficients and anomalous dimensions defined in the effective theory. 
Contributions associated with the physical scales $Q^2$ and $Q^2(1-x)$ are 
separated from non-perturbative hadronic physics in a transparent way. A 
crucial ingredient to the momentum-space resummation is the exact solution to 
the integro-differential evolution equation for the jet function, which is 
derived. The methods developed in this Letter can be applied to many other 
hard QCD processes.
\end{abstract}

\pacs{Preprint: CLNS~06/1963, FERMILAB-PUB-06-101-T}]
\narrowtext

\section{Introduction}

A generic problem in applications of perturbative QCD to collider physics or 
heavy-quark physics is to disentangle contributions associated with different 
momentum scales, and to resum large logarithms of ratios of such scales to all 
orders in the perturbative expansion. In processes containing hadronic jets, a 
scale hierarchy is created by the fact that the invariant mass of a collimated 
jet is typically much smaller than the hard scale of the process (e.g., the 
center-of-mass energy). The intricate interplay of soft and collinear 
emissions then leads to large Sudakov double logarithms. The resummation of 
these logarithms is conventionally performed in moment space, and predictions 
for differential cross sections in momentum space are obtained by an 
inverse Mellin transformation. This procedure is cumbersome and often leads to 
unphysical singularities, because the resummation formulae involve integrals 
over the Landau pole of the running coupling. These singularities are dealt 
with by means of ad hoc prescriptions, or by introducing artificial infrared 
cutoffs.

In this Letter we develop an approach based on effective field theory, which 
allows us to resum Sudakov logarithms for a large class of processes directly 
in momentum space. The starting point is a factorization theorem for the cross 
section, in which contributions from different momentum scales are separated 
in a transparent way. Evolution equations for the various components in the 
factorization formula are solved exactly in momentum space, in such a way that 
one never encounters integrals over the Landau pole. We illustrate the 
procedure with the example of deep-inelastic scattering. However, the same 
methods can be applied to many other hard QCD processes, such as Drell-Yan 
lepton-pair production, prompt photon production in hadron-hadron collisions, 
Higgs-boson production in gluon-gluon fusion, heavy-quark fragmentation, event shapes, and others. Technical details of our derivations are presented in \cite{LongPaper}.

\section{Factorization formula}

We focus on the flavor non-singlet component of the structure function 
$F_2(x,Q^2)$ in deep-inelastic scattering (DIS) of electrons off a nuclear 
target, $e^- +N(p)\to e^- +X(P)$, denoting by $q=P-p$ the momentum of the 
virtual photon. We are interested in the region where the Bjorken scaling 
variable $x=Q^2/(2p\cdot q)$ is near 1, so that there is a hierarchy of scales 
$Q^2\gg Q^2(1-x)\gg\Lambda_{\rm QCD}^2$. The intermediate scale 
$Q^2(1-x)\approx M_X^2$ is set by the invariant mass $M_X$ of the final-state
jet. In this region the structure function can be written in the factorized 
form \cite{Sterman:1986aj,Catani:1989ne,Korchemsky:1992xv} (with $\mu_f$ the 
factorization scale and $e_q$ the quark electric charge)
\begin{eqnarray}\label{fact}
   F_2^{\rm ns}(x,Q^2)
   &=& \sum_q\,e_q^2\,|C_V(Q^2,\mu_f)|^2 \nonumber\\[-0.1cm]
   &\times& Q^2\int_x^1\!d\xi\,J\!\left( Q^2(\xi-x),\mu_f \right)
    \phi_q^{\rm ns}(\xi,\mu_f) \,. \nonumber
\end{eqnarray}
This formula is valid to all orders in perturbation theory and at leading 
power in $(1-x)$ and $\Lambda_{\rm QCD}^2/M_X^2$. Here $C_V$ is a hard 
matching coefficient, $J$ is a jet function, and $\phi_q^{\rm ns}$ is the 
non-singlet component of the quark distribution function in the nucleon. As 
shown in \cite{LongPaper}, a simple derivation of the factorization 
formula can be given using the technology of soft-collinear effective theory 
(SCET) \cite{Bauer:2000yr} (see 
\cite{Manohar:2003vb,Pecjak:2005uh,Chay:2005rz} for earlier investigations). 

In SCET the hard function $C_V$ is identified with the Wilson
coefficient in the matching relation of the QCD vector current onto
the unique leading-power current operator in the effective theory. To
calculate the Wilson coefficient one must compare perturbative
expressions for the photon vertex function in the two theories. The
calculation can be simplified by performing the matching on-shell, in
which case all loop graphs in the effective theory are scaleless and
vanish in dimensional regularization. The bare on-shell vertex
function in QCD (called the on-shell quark form factor) has been
studied at two-loop order and beyond
\cite{Kramer:1986sg,Matsuura:1987wt,Moch:2005id}. The
form factor is infrared divergent and must be regularized. When the
SCET graphs are subtracted from the QCD result, the infrared poles in
$1/\epsilon$ get replaced by ultraviolet poles. To
obtain the matching coefficient we introduce a renormalization factor
$Z_V$, which absorbs these poles. At one-loop order this gives
\cite{Manohar:2003vb}
\[
   C_V(Q^2,\mu) = 1 + \frac{C_F\alpha_s}{4\pi}
   \left( - L^2 + 3L - 8 + \frac{\pi^2}{6} \right) ,
\]
where $L=\ln(Q^2/\mu^2)$ and $\alpha_s=\alpha_s(\mu)$. The two-loop expression 
for $C_V$ can be found in \cite{LongPaper}. The scale dependence of the Wilson 
coefficient is governed by the evolution equation
\begin{equation}\label{gammaV}
   \frac{dC_V(Q^2,\mu)}{d\ln\mu}
   = \left[ \Gamma_{\rm cusp}(\alpha_s)\,\ln\frac{Q^2}{\mu^2}
   + \gamma^V(\alpha_s) \right] C_V(Q^2,\mu) \,,
\end{equation}
where $\Gamma_{\rm cusp}$ is the universal cusp anomalous dimension of Wilson 
loops with light-like segments \cite{Korchemsky:1987wg}, which is associated 
with the appearance of Sudakov double logarithms. The quantity $\gamma^V$ 
accounts for single-logarithmic evolution effects. The anomalous dimension can 
be obtained from the coefficient of the $1/\epsilon$ pole term in the 
renormalization factor $Z_V$. Using the results of \cite{Moch:2005id} it can 
be calculated at three-loop order \cite{LongPaper}.

The jet function $J$ is defined in terms of the discontinuity of a vacuum 
correlator of two quark fields, made gauge invariant by the introduction of 
Wilson lines. It obeys the integro-differential evolution equation 
\cite{Becher:2006qw}
\begin{eqnarray}\label{Jrge}
   \frac{dJ(p^2,\mu)}{d\ln\mu}
   &=& - \left[ 2\Gamma_{\rm cusp}(\alpha_s)\,\ln\frac{p^2}{\mu^2}
    + 2\gamma^J(\alpha_s) \right] J(p^2,\mu) \nonumber\\
   &&\mbox{}- 2\Gamma_{\rm cusp}(\alpha_s) \int_0^{p^2}\!dp^{\prime 2}\,
    \frac{J(p^{\prime 2},\mu)-J(p^2,\mu)}{p^2-p^{\prime 2}} \,. \nonumber
\end{eqnarray} 
We encounter again the cusp anomalous dimension, and in addition a new 
function $\gamma^J$, which has been calculated in \cite{Becher:2006qw} at 
two-loop order, and whose three-loop coefficient is determined in 
\cite{LongPaper}.

\section{Solutions of the renormalization group equations}

The exact solution to the evolution equation (\ref{gammaV}) is
\begin{eqnarray}\label{CVsol}
   C_V(Q^2,\mu)
   &=& \exp\left[ 2S(\mu_h,\mu) - a_{\gamma^V}(\mu_h,\mu) \right] \nonumber\\
   &\times& \left( \frac{Q^2}{\mu_h^2} \right)^{-a_\Gamma(\mu_h,\mu)}\,
    C_V(Q^2,\mu_h) \,,
\end{eqnarray}
where $\mu_h\sim Q$ is a hard matching scale, at which the value of the  
coefficient $C_V$ is calculated using fixed-order perturbation theory. The 
Sudakov exponent $S$ and the exponents $a_\gamma$ are given by
\begin{eqnarray}\label{RGEsols}
   S(\nu,\mu)
   &=& - \int\limits_{\alpha_s(\nu)}^{\alpha_s(\mu)}\!
    d\alpha\,\frac{\Gamma_{\rm cusp}(\alpha)}{\beta(\alpha)}
    \int\limits_{\alpha_s(\nu)}^\alpha
    \frac{d\alpha'}{\beta(\alpha')} \,, \nonumber\\
   a_\Gamma(\nu,\mu) &=& - \int\limits_{\alpha_s(\nu)}^{\alpha_s(\mu)}\!
    d\alpha\,\frac{\Gamma_{\rm cusp}(\alpha)}{\beta(\alpha)} \,,
\end{eqnarray}
and similarly for $a_{\gamma^V}$, where $\beta(\alpha_s)=d\alpha_s/d\ln\mu$ is 
the $\beta$-function. The explicit perturbative expansions of these 
expressions valid at next-to-next-to-leading order (NNLO) in 
renormalization-group (RG) improved perturbation theory are given in 
\cite{LongPaper}.

An important object in the derivation of the solution to the evolution 
equation for $J$ is the associated jet function $\widetilde j$, which has 
originally been defined in terms of an integral over the jet function followed 
by a certain replacement rule \cite{Neubert:2005nt}. More elegantly, it can be 
obtained by the Laplace transformation
\[
   \widetilde j\Big( \ln\frac{Q^2}{\mu^2},\mu \Big)
   = \int_0^\infty\!dp^2\,e^{-s p^2}\,J(p^2,\mu) \,,
\]
where $s=1/(e^{\gamma_E} Q^2)$. The inverse transformation is
\begin{equation}\label{Mellin}
   J(p^2,\mu) = \frac{1}{2\pi i} \int_{c-i\infty}^{c+i\infty}\!ds\,e^{s p^2}\,
   \,\widetilde j\Big( \ln\frac{1}{e^{\gamma_E} s\,\mu^2},\mu \Big) \,.
\end{equation}
Using the evolution equation for the jet function we find that the associated 
jet function obeys
\begin{eqnarray}\label{jtildeevol}
   && \frac{d}{d\ln\mu}\,\widetilde j\Big( \ln\frac{Q^2}{\mu^2},\mu \Big)
    \nonumber\\
   &=& - \left[ 2\Gamma_{\rm cusp}(\alpha_s)\,\ln\frac{Q^2}{\mu^2}
    + 2\gamma^J(\alpha_s) \right]
    \widetilde j\Big( \ln\frac{Q^2}{\mu^2},\mu \Big) \,, \nonumber
\end{eqnarray}
which is analogous to the evolution equation (\ref{gammaV}) for the hard 
function. Inserting the solution to this equation into the inverse 
transformation (\ref{Mellin}) we obtain
\begin{eqnarray}\label{sonice}
   J(p^2,\mu)
   &=& \exp\left[ - 4S(\mu_i,\mu) + 2 a_{\gamma^J}(\mu_i,\mu) \right]
    \nonumber\\
   &\times& \widetilde j(\partial_\eta,\mu_i)\,
    \frac{e^{-\gamma_E\eta}}{\Gamma(\eta)}\,\frac{1}{p^2}
    \left( \frac{p^2}{\mu_i^2} \right)^\eta \,,
\end{eqnarray}
where $\eta=2a_\Gamma(\mu_i,\mu)$, and $\partial_\eta$ denotes a derivative 
with respect to this quantity. The above form of the result is valid as long 
as $\eta>0$ (i.e., $\mu<\mu_i$). For negative $\eta$ the singularity at 
$p^2=0$ must be regularized using a star distribution \cite{LongPaper}. 
Relation (\ref{sonice}) is one of the main results of this Letter. It relates 
$J$ to the associated jet function $\widetilde j$ evaluated at a scale 
$\mu_i$, where it can be computed using fixed-order perturbation theory. At 
one-loop order
\[
   \widetilde j(L,\mu) = 1 + \frac{C_F\alpha_s}{4\pi} 
   \left( 2L^2 - 3L +7 - \frac{2\pi^2}{3} \right) ,
\]
where in (\ref{sonice}) the argument $L$ is replaced by the derivative 
operator $\partial_\eta$. The two-loop expression for $\widetilde j$ can be 
extracted from \cite{Becher:2006qw}.

\section{Momentum-space resummation}

We are now ready to write down a resummed expression for the structure 
function $F_2^{\rm ns}(x,Q^2)$ valid to all orders in perturbation theory and 
at leading power in $(1-x)$ and $\Lambda_{\rm QCD}^2/M_X^2$. When combining 
the results (\ref{CVsol}) and (\ref{sonice}) the Sudakov exponents can be 
simplified. Introducing the short-hand notation 
$a_{\gamma^\phi}=a_{\gamma^J}-a_{\gamma^V}$, we find after a straightforward 
calculation
\begin{eqnarray}
   F_2^{\rm ns}(x,Q^2)
   &=& \sum_q e_q^2\,|C_V(Q^2,\mu_h)|^2\,U(Q,\mu_h,\mu_i,\mu_f) \nonumber\\
   &&\hspace{-1.3cm}
    \times \widetilde j\Big( \ln\frac{Q^2}{\mu_i^2}+\partial_\eta,\mu_i
    \Big)\,\frac{e^{-\gamma_E\eta}}{\Gamma(\eta)}\,\int_x^1\!d\xi\,
    \frac{\phi_q^{\rm ns}(\xi,\mu_f)}{\left( \xi-x \right)^{1-\eta}} \,,
\label{wonder}\end{eqnarray}
where 
\begin{eqnarray}
   U(Q,\mu_h,\mu_i,\mu_f)
   &=& \exp\left[ 4S(\mu_h,\mu_i) - 2a_{\gamma_V}(\mu_h,\mu_i) \right]
    \nonumber\\
   &&\hspace{-2.1cm}
    \times \left( \frac{Q^2}{\mu_h^2} \right)^{-2a_\Gamma(\mu_h,\mu_i)} 
    \exp\left[ 2a_{\gamma^\phi}(\mu_i,\mu_f) \right] \,, \nonumber
\end{eqnarray}
and as before $\eta=2a_\Gamma(\mu_i,\mu_f)$. The remaining integral can be 
performed noting that, on general grounds, the behavior of the parton 
distribution function near the endpoint can be parameterized as
\[
   \phi_q^{\rm ns}(\xi,\mu_f) \big|_{\xi\to 1}
   = N(\mu_f)\,(1-\xi)^{b(\mu_f)} \Big[ 1 + {\cal O}(1-\xi) \Big] \,,
\]
where $b(\mu_f)>0$. This leads to the final expression
\begin{eqnarray}\label{beauty}
   \frac{F_2^{\rm ns}(x,Q^2)}{\sum_q e_q^2\,x\,\phi_q^{\rm ns}(x,\mu_f)}
   &=& |C_V(Q^2,\mu_h)|^2\,U(Q,\mu_h,\mu_i,\mu_f) \nonumber\\
   &&\hspace{-2.3cm}
    \times (1-x)^\eta\,\,
    \widetilde j\Big( \ln\frac{Q^2(1-x)}{\mu_i^2}+\partial_\eta,\mu_i \Big)
    \nonumber\\
   &&\hspace{-2.3cm} \times
    \frac{e^{-\gamma_E\eta}\,\Gamma(1+b(\mu_f))}{\Gamma(1+b(\mu_f)+\eta)} \,.
\end{eqnarray}
The exact all-order results (\ref{wonder}) and (\ref{beauty}) are independent of the
scales $\mu_h$ and $\mu_i$, at which the matching coefficient $C_V$
and the associated jet function $\widetilde j$ are calculated. The
answers simplify further if we choose the ``natural'' values
$\mu_h=Q$ and $\mu_i=Q\sqrt{1-x}$ (for fixed $x$). In practical
calculations the residual dependence on these scales introduced
by the truncation of the perturbative expansions of the various
objects can be used as an estimator of yet unknown higher-order
corrections.

Above we have accomplished the resummation of threshold logarithms for
$F_2$ directly in momentum space. The resulting formulae are simpler
than corresponding expressions in the literature (see e.g.\
\cite{Moch:2005ba}) in that they do not require a Mellin inversion and
in that the dependence on $x$ and $Q$ is explicit. The right-hand
sides of (\ref{wonder}) and (\ref{beauty}) can be evaluated at any
desired order in resummed perturbation theory. Using currently
available results, it is possible to include terms at NNLO
\cite{LongPaper}, which is equivalent to the so-called
next-to-next-to-next-to-leading double-logarithmic (N$^3$LL)
approximation. The resummation is under perturbative control as long
as $(1-x)\gg\Lambda_{\rm QCD}^2/Q^2$, since only then the intermediate
scale $\mu_i\sim Q\sqrt{1-x}$ is a short-distance scale. While the
theoretical description thus breaks down very close to the endpoint,
we note that weighted integrals of the structure function over an
interval $x_0\le x\le 1$ can be calculated as long as $Q\sqrt{1-x_0}$
is in the short-distance domain.

It is instructive to compare our result (\ref{beauty}) with the conventional 
approach to threshold resummation in DIS, which proceeds via moment space 
\cite{Sterman:1986aj,Catani:1989ne}. One defines
\begin{eqnarray}
   F_{2,N}^{\rm ns}(Q^2) 
   &=& \int_0^1\!dx\,x^{N-1} F_2^{\rm ns}(x,Q^2) \nonumber\\
   &=& C_N(Q^2,\mu_f)\,\sum_q\,e_q^2\,\phi_{q,N+1}^{\rm ns}(\mu_f) \,, \nonumber
\end{eqnarray}
where the moments of $\phi_q^{\rm ns}(\xi,\mu)$ are defined in analogy with those of $F_2^{\rm ns}(x,Q^2)$. For large values of $N$ the integral is dominated by the endpoint region 
$(1-x)\sim 1/N$. The short-distance coefficient $C_N$ is decomposed as
\[
   C_N(Q^2,\mu_f) = g_0(Q^2,\mu_f)\,\exp\left[ G_N(Q^2,\mu_f) \right] ,
\]
where the prefactor $g_0$ collects all $N$-independent terms, and the exponent 
is written in the form (see \cite{Moch:2005ba} for the most up-to-date 
discussion)
\begin{eqnarray}\label{GN}
   G_N(Q^2,\mu_f)
   &=& \int_0^1\!dz\,\frac{z^{N-1}-1}{1-z} \\
   &&\hspace{-1.8cm}\times
    \left[ \int_{\mu_f^2}^{(1-z)Q^2}\!\frac{dk^2}{k^2}\,A_q(\alpha_s(k))
    + B_q\!\left( \alpha_s(Q\sqrt{1-z}) \right) \right] . \nonumber
\end{eqnarray}
The resummation for the momentum-space structure function $F_2(x,Q^2)$ itself 
is obtained from that for the moments $F_{2,N}(Q^2)$ by an inverse Mellin 
transformation. It is possible to show (see \cite{LongPaper} for details) that 
the outcome of this procedure is equivalent, at any finite order in the 
perturbative expansion, to the result (\ref{beauty}) derived from effective 
field theory, provided we identify $A_q(\alpha_s)=\Gamma_{\rm cusp}(\alpha_s)$ 
and 
\begin{eqnarray}
   \left( 1 + \frac{\pi^2}{12}\,\nabla^2 + \dots \right) B_q(\alpha_s)
   &=& \gamma^J(\alpha_s) + \nabla\,\ln\widetilde j(0,\mu) \nonumber\\
   &&\hspace{-3.3cm}\mbox{}- \left( \frac{\pi^2}{12}\,\nabla
    - \frac{\zeta_3}{3}\,\nabla^2 + \dots \right) \Gamma_{\rm cusp}(\alpha_s)
    \,, \nonumber
\end{eqnarray}
where $\nabla=d/d\ln\mu^2$. It follows from this relation that the quantities 
$B_q$ and $\gamma^J$ agree at first order in $\alpha_s$ (as observed in 
\cite{Manohar:2003vb}), but they differ starting from two-loop order. 

There are a few unpleasant features of the conventional approach which
are worth pointing out. First, note that the integrals over the
functions $A_q$ and $B_q$ in (\ref{GN}) run over the Landau pole of
the running coupling $\alpha_s(\mu)$, introducing an infrared
renormalon ambiguity of order $\Lambda_{\rm QCD}^2/M_X^2$. No such
problem arises for the integrals (\ref{RGEsols}) in our approach. The
particular integral representation of the solution (\ref{GN}) results
from the fact that in the conventional approach one solves a set of
partial differential equations derived by diagramatic methods instead
of the RG evolution equations in SCET
\cite{Sterman:1986aj,Catani:1989ne}. The use of RG methods for the
resummation avoids the Landau-pole singularity in the exponent
\cite{Korchemsky:1993uz,Beneke:1995pq}.  Secondly, when performing the
inverse Mellin transform one needs to integrate the function $G_N$
over $N$ along a contour parallel to the imaginary axis. This
integration involves arbitrarily small physical scales $|k^2|\sim
Q^2/|N|$, leading to a second encounter with the Landau
pole. Different prescriptions to deal with this problem have been
proposed in the literature. In our approach integrals over the Landau
pole never arise, because factorization and resummation are 
performed directly in momentum space. The singularities appearing in the
conventional approach are an artifact of the way the resummation of
large logarithms is implemented and they cannot be used to assess
whether a corresponding renormalon pole is present
\cite{Beneke:1995pq}. In our approach infrared renormalons are
expected to affect the large-order perturbative behavior of the
matching coefficients $C_V$ and $\tilde{j}$. The corresponding
infrared ambiguities will be commensurate with power corrections from
subleading operators in the effective theory. The evolution, on the
other hand, is driven by anomalous dimensions, which are expected to
be free of renormalons.

\begin{figure}
\epsfxsize=8.7cm
\centerline{\epsffile{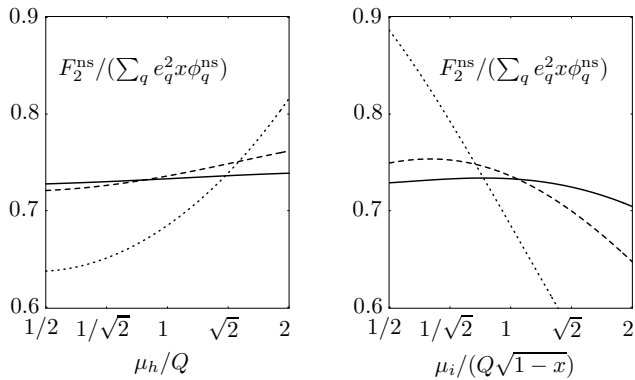}}
\vspace{0.2cm}
\centerline{\parbox{14cm}{\caption{\label{fig:plot}
Dependence of the resummed result for $F_2^{\rm ns}(x,Q^2)$ on the hard (left) 
and intermediate (right) scales, at different orders in RG-improved 
perturbation theory: LO (dotted), NLO (dashed), and NNLO (solid). We use 
$Q=30$\,GeV, $x=0.9$, $\mu_f=5$\,GeV, and $b(\mu_f)=4$.}}}
\end{figure}

Figure~\ref{fig:plot} shows the scale dependence of our result for 
$F_2^{\rm ns}$ obtained with $Q=30$\,GeV, $x=0.9$, $\mu_f=5$\,GeV, and $n_f=5$ 
light flavors. Varying the scales $\mu_h$ and $\mu_i$ about their 
default values, we observe that the residual scale dependence is strongly 
reduced when going to successively higher orders in perturbation theory. In 
the literature the matching scales are often held fixed, and the perturbative 
uncertainties can only be estimated by comparing results at different orders in the 
expansion.

\section{Conclusions}

Using methods from effective field theory we have introduced a new
approach to the resummation of large Sudakov logarithms in hard QCD
processes. Factorization and resummation are performed directly in
momentum space, such that the resulting formulae are free of
unphysical infrared sensitivities and provide a transparent separation
of the different scales in the problem. As an application we have
derived an exact all-order expression for the resummed deep-inelastic
structure function $F_2$, which is much simpler than corresponding
results found in the literature.

\vspace{0.3cm}  
{\em Acknowledgments:\/} 
We are grateful to Ben Pecjak for useful discussions. The research of T.B.\ 
was supported by the Department of Energy under Grant DE-AC02-76CH03000, and 
that of M.N.\ by the National Science Foundation under Grant PHY-0355005. 
Fermilab is operated by Universities Research Association Inc., under contract 
with the U.S.\ Department of Energy.

\end{document}